\documentclass[prb,12pt]{revtex4}

\usepackage[english]{babel}
\usepackage[intlimits]{amsmath}
\usepackage[dvips]{graphicx}

\begin{document}

\begin{titlepage}
\
\title{Polymerization dynamics of double-stranded biopolymers:  chemical  kinetic approach}
\author{Evgeny B. Stukalin and Anatoly B. Kolomeisky}

\affiliation{Department of Chemistry, Rice University, Houston, TX 77005 USA}

\begin{abstract}

The polymerization dynamics of double-stranded polymers, such as actin filaments, is investigated theoretically  using simple chemical kinetic models that explicitly take into account  some microscopic details of the polymer structure and the lateral interactions between the protofilaments. By considering all possible molecular configurations, the exact analytical expressions for the growth velocity and dispersion for  two-stranded polymers are obtained in the case of the growing  at only one end, and for the  growth from both polymer ends.  Exact theoretical calculations are compared with the predictions of approximate multi-layer models that consider only a finite number of the most relevant polymer configurations. Our theoretical approach is applied to analyze the experimental data on the growth and fluctuations dynamics of individual single actin filaments.  

\end{abstract}

\maketitle

\end{titlepage}

\section {Introduction}

Cytoskeletal   proteins such as actin filaments, intermediate filaments and microtubules are rigid multifilament polymers that play a variety of roles in biological systems, including organization of cell structures, transport of organelles and vesicles, cell motility and reproduction.\cite{howard_book,bray_book,pollard03} Biological functions of these proteins are mostly determined by  the processes that take place during their polymerization. However, our understanding  of the coupling between these biopolymer's structure and  functions and their growth processes is still very limited.     

In recent years the number of experimental investigations of the growth mechanisms and dynamic properties of  rigid multifilament biopolymers  at a single-molecule level increased significantly.\cite{erickson92,desai97,dogterom97,shaw03,dogterom03,janson04,lehto03,fujiwara02} Dynamic behavior of individual microtubules have been characterized by a variety of experimental techniques such as video and  electron microscopy, fluorescence spectroscopy, and optical trap spectrometry,\cite{erickson92, desai97,dogterom97,shaw03,dogterom03,janson04} whereas the studies of the single actin filaments have just begun.\cite{fujiwara02, lehto03} Many unusual phenomena in the assembly dynamics of these  biopolymers  have been observed, such as  treadmilling for microtubules and actin filaments,\cite{fujiwara02, hotani88, grego01} and microtubule dynamic instability. \cite{hotani88,mitchison92} 

Recent experimental investigations of the single actin filament length fluctuations\cite{fujiwara02} raised many questions about  the actin polymerization dynamics. A large discrepancy  in kinetic rate constants estimated by average length change in the initial polymerization phase and  from the analysis of length fluctuations in the steady-state phase (a factor of  40) has been observed. One of the possible explanations for this discrepancy might be the oversimplified theoretical model, used in analysis,\cite{fujiwara02} that neglected the polymer structure and  lateral monomer-monomer interactions at the actin filament tips.  Similar problems have been found in the growth dynamics of individual microtubules under the influence of external forces.\cite{dogterom97,KF01,janson04} For example, the depolymerization rate constants determined from a phenomenological description of microtubule dynamics, that do not incorporate the biopolymer's structure and lateral interactions between the filaments, contradict to values measured in some independent bulk chemical kinetic experiments.\cite{KF01} 

A large volume of experimental results stimulated many theoretical investigations of polymerization dynamics for rigid multifilament proteins. In one approach, it was suggested that the growth of rigid biopolymers is controlled by thermal fluctuations. \cite{peskin93,mogilner99,van Doorn00} This is a basic idea of the so-called polymerization ratchet models. In a different approach, more phenomenological chemical kinetic (stochastic) models have been used to describe the biopolymer's growth dynamics.\cite{hill86, freed02, limbach99, ermentrout00}  In these simplified  phenomenological models it is assumed that the overall dynamics is a balance between polymerization and depolymerization processes, however the microscopic details  of the polymer structure and  the differences in lateral inter-subunit interactions are not taken into consideration. In addition, the microtubule assembly dynamics has been studied extensively by computer simulations. \cite{chen85,bayley90,martin93,vanburen02} 

Recently, we introduced a set of simple stochastic models for the description of the growth  of rigid biopolymers  consisting of $N$ protofilaments, that explicitly includes the geometric structure and  monomer-monomer lateral interactions.\cite{stukalin04} According to this approach,  only few polymer configurations are relevant for the  growth dynamics  because of inhomogeneity in lateral interactions that comes from the geometry of  polymer ends. These configurations are selected using the criteria that the distances between the protofilament tips in each configuration should be less than $n d$, where $n=1,2,\cdots$ and $d$ is a polymer subunit length. For  $n=1$, in the so-called ``one-layer'' approximation, the mean growth velocity and a dispersion (or a diffusion coefficient) of polymer's length have been calculated exactly for any number of protofilaments and for the arbitrary geometry of biopolymer's growing end. The quality of the ``one-layer''  approximation has been discussed for the simple case of the growth of polymer made of two protofilaments, i.e., $N=2$, which is closely related to  actin filaments. In this case, a full dynamic description that includes all possible polymer configurations  provided exact expressions for the mean growth velocity. It was found that the predictions of the approximate model are approaching the exact results for large  but realistic values of the lateral interactions.

Although the approach presented in Ref.\cite{stukalin04} provided a very good  description of experimental results  of individual  microtubule growth under external forces, and it suggested a reasonable way  of  coupling of the microscopic structure of the biopolymers  with their dynamic properties, there are several theoretical problems with this  method. First, the criteria that helps to determine the finite number of the most relevant configurations in ``$n$-layer'' approximations is introduced  using an arbitrary cut-off distance, but it would be more desirable to  derive it from more fundamental thermodynamic and kinetic arguments. Second, the method used for calculating exactly the full dynamic model of the growth of two-stranded polymers allowed only the determination of the mean growth velocity, and not the dispersion. However, the simultaneous knowledge of the dispersion and the velocity is crucial for understanding the  growth mechanisms of rigid biopolymers.\cite{howard_book,KF01} The goal of this paper is to address these theoretical issues  by studying the growth dynamics of two-stranded biopolymers with more detailed microscopic description. In addition, we will analyze the experimental observations on the growth of single actin filaments. 

This article is organized as follows. The dynamics of the two-stranded polymers that can grow only from one end is discussed in Section II, while the description of the polymer growth from  both ends is given in Section III. The application of the developed chemical kinetic models for the  experiments on single actin filaments is presented in Section IV. The results are discussed and summarized in Section V. The mathematical details of calculations of dynamic properties are given in Appendix.

\section{Growth dynamics of attached two-stranded polymers}

Consider a growing rigid polymer that consists of two protofilaments. The polymer is attached to a nucleating site on a surface and may polymerize or depolymerize only from  the  free end, as shown in Fig. 1. This is the attached two-stranded polymer. \cite{hill} The building block of this polymer is a monomer subunit of  length $d$. The lateral offset between two parallel protofilaments  is equal to $a$ and it can generally vary between 0 and $d$. For actin filaments the monomer size is $d=5.4$ nm and the lateral shift is $a=d/2=2.7$ nm.\cite{howard_book,bray_book}

There are  infinite number of possible polymer configurations that differ from each other by the geometry of the growing end and the total length of the polymer. For labeling these configurations we use a pair of integers that count the number of monomer subunits in each protofilament. Without loss a generality, let us choose a first (bottom) protofilament as a leading one in a configuration $(l,l)$, where the distance between protofilament tips  equal to $a$: see Fig. 1. The second (top) protofilament will be the leading one in the configuration labeled as $(l,l+1)$, where the distance between the tips is equal to $d-a$. Generally, as illustrated in Fig. 1, for configurations $(l,l-k)$, where $k=0,1,2,\cdots$, the distances between the parallel protofilaments are equal to $a+kd$ and the first protofilament is the leading one, while for configurations $(l-k,l+1)$ the distances between  protofilament tips are $d-a+kd$ and the second protofilament is  the leading one. 

All dynamic transitions in the attached two-stranded polymers  may be characterized by a set of four pairs of transition rates that depend on the local geometry of the growing end.  As shown in Fig. 1, the attachment and detachment rates from the leading protofilament for all polymer configurations are given by $u_0$ and $w_0$, respectively. Transition rates $u_{1}$ and $w_{1}$ describe the polymerization and depolymerization events when the overall length of the polymer does not change. For example, as presented in Fig. 1, the transitions between the configurations $(l+1,l)$ and $(l+1,l+1)$ are given by these rates, while in both configurations the polymer length is $d(l+1)+a$. A monomer can attach with the rate $u_{\delta}$ to the protofilament 2 of the configuration $(l,l)$, or it can detach with the rate $w_{\delta}$ from the same protofilament of the configuration $(l,l+1)$. Similarly, the monomer subunit can associate with the rate $u_{1-\delta}$ to the protofilament 1 of the configuration $(l,l+1)$, or it can dissociate with the rate $w_{1-\delta}$ from the same protofilament of the configuration $(l+1,l+1)$: see Fig. 1. Note that the subscript indexes (0, 1, $\delta$ and $1-\delta$, where $\delta=a/d$) are equal to the fractions of the lateral bond between the monomers on the parallel protofilaments created or broken in a given transition.

The overall kinetic scheme of the system, that includes all possible states and transitions, is shown in Fig. 2a. The  polymer growth dynamics can be described by a set of master equations for each configuration. Solutions of these equations, that characterize the steady-state growth dynamics of an attached two-stranded polymer, are outlined in a full detail in the Appendix A. Below we present only the final results for the mean growth velocity and dispersion. 

Within  the full dynamic description, the growth velocity appears formally as a sum of two terms, namely,
\begin{equation}\label{vel1}
V = V_0(1 - \beta) + V_1, 
\end{equation}
where
\begin{equation}\label{vel2}
V_0  = d\frac{{u_\delta  u_{1 - \delta }  - w_\delta  w_{1 - \delta } }}
{{u_\delta   + w_\delta   + u_{1 - \delta }  + w_{1 - \delta } }},
\end{equation}
and
\begin{equation}\label{vel3}
V_1  = d(u_0  - w_0 \beta ).
\end{equation}
The parameter $\beta$ ($0 < \beta < 1$) is given by
\begin{equation}
\beta  = \frac{{u_0  + w_1 }}{{u_1  + w_0 }}.
\end{equation}

The expression for the diffusion coefficient is also consists of two terms,
\begin{equation}\label{dif1}
D= D_0(1 - \beta) + D_1, 
\end{equation}
where the terms are given by the following expressions,
\begin{equation}\label{dif2}
D_0  = \frac{{d^2 }}{2} \frac{{(u_\delta  u_{1 - \delta }  + w_\delta  w_{1 - \delta }  - 2 A_0^2 )}} {{u_\delta   + w_\delta   + u_{1 - \delta }  + w_{1 - \delta } }}, 
\end{equation}
and
\begin{equation}\label{dif3}
 D_1  = \frac{{d^2 }}{2}\left[ u_0  + w_0 \beta  - \frac{{2( A_0  + w_0 )(u_0  -  A_0 \beta )}}{{u_1  + w_0 }} \right].
\end{equation}
The auxiliary function $A_{0}$ is defined as
\begin{equation}
A_0  = V_0/d.
\end{equation}

The dynamic properties of the growing polymer strongly depend on the lateral interactions between the parallel protofilaments. It can be seen from the fact that the transition rates for binding or unbinding the monomer subunit are directly related to a lateral interaction free energy  per monomer $g_h$ via the detail balance conditions. It can be shown that\cite{stukalin04}
\begin{equation}\label{db}
u_s /w_s  = u_0 /w_0 \gamma ^{2s} , \hspace{1cm} s = 0, \delta, 1 - \delta, 1,
\end{equation}
where 
\begin{equation}\label{gamma}
\gamma  = \exp ( - g_h /k_B T).
\end{equation}

This observation is the basis for the approximate theoretical description of the growth of rigid biopolymers with $N$ filaments, the one-layer model, that we developed earlier.\cite{stukalin04} In this model, only polymer configurations with the distances between the protofilament tips  less than the monomer length $d$ are considered. It allows then to calculate the  mean growth velocity and dispersion explicitly. Specifically, for $N=2$ it can be shown that
\begin{equation}
V_{one-layer} = V_0, \quad \mbox{and} \quad  D_{one-layer}= D_0.  
\end{equation}

Since the full dynamic description of the growth of the attached two-stranded polymer is now available, the quality of the one-layer approximation can be easily checked in this case. As shown in Fig. 3, the one-layer model describes the growth dynamics reasonably well for large lateral interactions. In the limit of infinite lateral interactions the predictions from the approximate theory become exact. However, the convergence of the approximate one-layer results to exact quantities strongly depends on the geometry of the growing polymer end, specifically, on  the ratio between the lateral shift $a$  and the monomer length $d$. The best description can be obtained for the symmetric case $\delta=a/d=1/2$, while for other geometries this approximation is less successful.    

One of the advantages of the one-layer approximation is its ability to be easily extended to include more polymer configurations. As a better approximation, it is natural to consider configurations where the distances between the protofilament tips do not exceed 2$d$, i.e., a two-layer model. The number of polymer configurations is still finite and the dynamic properties can be easily calculated following the approach presented for the one-layer approximation.\cite{stukalin04} In the two-layer approximation, for two-stranded polymers it can be shown that\begin{equation}
V_{two-layer}=\frac{1}{1+\beta}(V_0 + V_1),
\end{equation}
and
\begin{equation}
D_{two-layer}=\frac{1}{1+\beta}(D_0 + \widetilde D_1),
\end{equation}
where 
\begin{equation}
\widetilde D_1= \frac{{d^2 }}{2}\left[ u_0  + w_0 \beta  - \frac{{2( \widetilde A  + w_0 )(u_0  - \widetilde  A \beta )}}{{u_1  + w_0 }} \right],
\end{equation}
with the parameter $\widetilde A$ given by
\begin{equation}
\widetilde A = V_{two-layer}/d.
\end{equation}
The results of the two-layer approximation for the mean growth velocity and dispersion are also presented in Fig 3. The agreement with exact full dynamic properties is very good even for weak lateral interactions ($g_{h}>2 k_{B}T$), and the results do not depend much on the specific geometry of the growing polymer end.   

The multi-layer  approach can be used to describe the growth dynamics of any rigid polymer consisting of $N$ parallel protofilaments. It is important to understand thermodynamic and kinetic justifications for this approximations. The fact that the growth dynamics for two-stranded polymers ($N=2$) can be analyzed exactly is very useful to make the connection with $n$-layer approximate description. The kinetic schemes for full dynamic description and for the one-layer and two-layer approximations for two-stranded polymers are shown in Fig. 2. Comparing different kinetic diagrams, we can see that the one-layer approach corresponds to the main chemical pathway, while the two-layer approximation also takes into account the closest branched states. Thus, the $n$-layer approximations can be thought of as a series expansion (with $n$ terms) of full dynamic description, where the value of energy of lateral interactions determines how good is the expansion. The higher the lateral interactions the smaller number of terms is needed in order to describe successfully the growth dynamics of rigid multi-filament polymers.

\section{Growth dynamics of  free two-stranded polymers}

Now consider a two-stranded polymer that can freely grow from both ends. Define $x_{L}(t)$ and $x_{R}(t)$ as time-dependent coordinates of the ``left'' and ``right'' ends of the polymer. The  growth velocity is defined as
\begin{equation}
V_{free} = \frac{d}{dt}   \langle [x_{R} (t) - x_{L} (t)] \rangle, 
\end{equation}
and the angular brackets mean averaging  over all possible growth pathways. It can be easily seen that the mean growth velocity of a free polymer is a difference between two terms,
\begin{equation}
V_{free} = \frac{d}{dt} \langle x_{R}(t) \rangle  - \frac{d}{dt} \langle x_{L}(t) \rangle= V_{R}-V_{L},
\end{equation}
where $V_{R}$ and $V_{L}$ are one-end growth velocities, determined explicitly in Eqs. (\ref{vel1}), (\ref{vel2}), and  (\ref{vel3}).
 
Similarly, the expression for the dispersion is given by
\begin{equation}
D_{free} = \frac{1}{2}\frac{d}{{dt}}\left[ {\left\langle {[x_R (t) - x_L (t)]^2 } \right\rangle  - \left\langle { [x_R (t) - x_L(t)] } \right\rangle ^2 } \right].
\end{equation}
It can be simplified into the following equation,
\begin{equation}
D_{free} = \frac{1}{2}\frac{d}{{dt}}\left[ \langle x_R(t)^{2} \rangle  + \langle x_L (t)^2 \rangle  - \langle x_{R}(t) \rangle^{2} - \langle x_L(t)\rangle^{2} -2 \langle x_{R}(t) x_{L}(t) \rangle + 2 \langle x_{R}(t) \rangle \langle x_{L}(t) \rangle  \right].
\end{equation}
The polymerization dynamics at both ends are independent from each other, that means that 
\begin{equation}
\langle x_{R}(t) x_{L}(t) \rangle = \langle x_{R}(t) \rangle \langle x_{L}(t) \rangle. 
\end{equation}
This leads to the conclusion that the dispersion of free growing polymer can  be presented as a sum of two one-end dispersion terms,
\begin{equation}
D_{free} = D_{R}+D_{L},
\end{equation}
where the explicit expressions for $D_{R}$ and $D_{L}$ are given by Eqs. (\ref{dif1}), (\ref{dif2}) and (\ref{dif3}).

\section{Application of chemical kinetic models for the description of experiments on individual actin filaments}

The growth  dynamics of the single actin filaments has been studied experimentally using the fluorescence microscopy with total internal reflection.\cite{fujiwara02} The assembly of actin filaments was observed during the "polymerization" phase, i.e. at initial stages of the process, as well as the steady-state conditions, when the growing rate of the barbed end was  compensated  by the shortening at the pointed end. The main finding was that the kinetic rate constants estimated from the length change in the single actin filaments for the initial  period differ considerably from that estimated using the length fluctuation analysis in the  steady-state phase. The set of rate constants measured at the initial polymerization conditions mainly agrees with values obtained in other experimental studies,\cite{pollard00, carlier97, selden02}  while the rate constants estimated from the measurements of length fluctuations at steady-state conditions were 30-45 times higher. 

Several possible explanations for this discrepancy has been suggested.\cite{fujiwara02} The first one is that the kinetic constants obtained in the steady-state phase may be intrinsically different from those obtained in the initial phase of polymerization. The change of nucleotide composition of the growing, or shortening tips in the time course of the process was indicated as a probable cause. The depolymerization velocity of ADP-actin is known to be an order of magnitude higher than that of ATP-actin at the barbed end.  The second possible reason for the discrepancy in the rate constants may be due to the possibility that the "effective" size of polymerization-depolymerization unit may not necessarily correspond to a monomer. The authors speculate that one plausible way to eliminate this divergence is to set "effective" size of unit 5-6 times higher. However, this contradicts to widely accepted picture that the elementary step in the growth of actin filaments is adding or removing a single actin monomer.\cite{pollard00, carlier97, selden02}

In analyzing the experimental data on growth dynamics of single actin filaments the simplified phenomenological picture has been used.\cite{fujiwara02} Here, we investigate another possibility to explain the difference in the kinetic rate constants by using a chemical kinetic model with better description of polymer ends geometry and chemical interactions between monomers.

To estimate the parameters that describe the growth dynamics of actin filaments we note that $\delta=a/d=1/2$, and, using the detailed balance conditions [see Eqs. (\ref{db},\ref{gamma})], the rate constants can be written in the following form
\begin{eqnarray}\label{actin_rates}
 u_{\delta} = u_{1-\delta}= u_0 \gamma^{f_{1/2}+1/2}, & \quad  u_1 = u_0 \gamma^{f_1+1}, \nonumber \\
 w_{\delta} = w_{1-\delta}= w_0 \gamma^{f_{1/2}-1/2}, &  \quad w_1 = w_0 \gamma^{f_1-1}. 
\end{eqnarray}
Coefficients $f_{1/2}$ and $f_1$ reflect the different values of activation energies for specific polymerization and depolymerization events. Although the exact values of these parameters cannot be measured experimentally, they might be estimated quite realistically as $-0.5 \le f_{1/2} \le 0.5$ and $-1 \le f_1 \le 1$. It implies that the subunit attaches faster to the site where the stronger lateral contact is created. Similarly, the detachment is slower if a stronger lateral bond should be broken. For simplicity, in our calculations we consider only the case
\begin{equation} 
f_{1/2} = f_1 = 0, 
\end{equation}
 and, as we checked, for other values of these parameters the results do not deviate much from the one presented here. 

Equations (\ref{actin_rates}) imply that the growth dynamics of actin filaments can be described by using only 3 parameters: $u_{0}$, $w_{0}$ and $\gamma$. The parameters  $u_{0}=k_{0} C$ (where $C$ is the concentration of free actin monomers in the solution) and $w_{0}$ are the association and dissociation rates from the leading protofilaments, and  $\gamma$ is a measure of lateral interactions in actin filaments. The mean growth velocity for each end of actin filaments can be presented in a simple form,
\begin{equation}\label{velocity}
V=\frac{d}{2}(u_{0}-w_{0}/\gamma)(\gamma^{1/2}-\gamma^{-1/2}+2),
\end{equation}
while the dispersion of the polymer length at each end  is given by more complex expression,
\begin{equation}\label{dispersion}
D=\frac{d^{2}}{2} \left[ \frac{1}{4}(u_{0}+w_{0}/\gamma)(\gamma^{1/2}+4-5\gamma^{-1/2}+2/\gamma)-(1-2 \gamma^{-1/2}+1/\gamma) \frac{2u_{0}w_{0}/\gamma}{u_{0}+w_{0}/\gamma} \right].
\end{equation}

 In order to apply our explicit expressions to describe the single actin filaments growth the  elongation rate constants for each end of the polymer should be known. However in the single-molecule experiments by Fujiwara et al.\cite{fujiwara02} the growth dynamics of each end separately has not been measured.  Nevertheless, for calculations we can use the data from other investigations  where the polymerization dynamics at both ends has been characterized quantitatively in the similar experimental conditions.\cite{pollard00} Kinetics of actin polymerization for the  barbed end  can be  described as $V^{b}=k_{+}^{b}C-k_{-}^{b}$ with $k_{+}^{b}=11.6$ $\mu$M$^{-1}$s$^{-1}$ and $k_{-}^{b}=1.4$ s$^{-1}$. Comparing this phenomenological expression with the exact one [see Eq. (\ref{velocity})] allows us to estimate the parameters $u_{0}$ and $w_{0}$. Here we also use the realistic estimate of energy of lateral interactions $g_h \sim 6$ $k_{B}T$ \cite{erickson89} that gives $\gamma \simeq 400$. These parameters are then applied to compute the contribution to the dispersion from the barbed end using  Eq. (\ref{dispersion}). Similar approach is utilized for the pointed end, for which the mean growth velocity  can be  described phenomenologically as  $V^{p}=k_{+}^{p}C-k_{-}^{p}$ with $k_{+}^{p}=1.3$ $\mu$M$^{-1}$s$^{-1}$ and $k_{-}^{p}=0.8$ s$^{-1}$.\cite{pollard00}  As a result, the overall dispersion of single actin filaments at steady-state concentration $C_{0}=0.17$ $\mu$M can be estimated as  $D \simeq 1.0 \times 10^{-3}$ $\mu$m$^{2}$/min. It should be noted that this procedure depends weakly on the value of $\gamma$. Also in these calculations we used the subunit length $d=5.4$ nm, and lateral off-set $a=2.7$ nm.

In the single actin filaments experiments\cite{fujiwara02} the measurement of fluctuations at steady state conditions produced the dispersion of $D \simeq 1.1 \times 10^{-2}$ $\mu$m$^{2}$/min, which is approximately 10 times larger than the value calculated above. The difference is significant and it implies that the chemical kinetic models with detailed description of polymer ends and monomer-monomer interactions still cannot explain fully the experimentally observed fluctuations in growing actin filaments. However, our theoretical treatment does not take into account the hydrolysis of actin-ATP monomers and related processes. It might be expected that these processes can significantly effect the growth dynamics of actin filaments.

It is interesting to note that the dispersion for actin filament assembly (with $a=d/2$) is a non-linear function of monomer's concentration, as shown in Fig. 4. At high concentrations of actin monomers the dispersion is proportional to concentration, while for low concentrations there is a weak deviation from  linearity. This dependence contrasts to the observed and calculated behavior of the mean growth velocity. It will be interesting to measure experimentally the concentration dependence of dispersion.

\section{Summary and conclusions}

We investigated theoretically the growth dynamics of two-stranded polymers where association and dissociation of monomers can take place from both ends. Because the polymerization events at each end are independent from each other, we argued that the overall polymer elongation dynamics can be described as a combination of growth processes at each end separately. 

For attached rigid two-stranded polymers, that  made of two protofilaments and can only elongate from one end, we developed a chemical kinetic model of the growth. This model takes into account the exact relative positions of two protofilaments and both lateral and longitudinal chemical interactions between the monomers. Considering full dynamic chemical kinetic scheme, the exact and explicit expressions for the mean growth velocity and dispersion have been derived in terms of rate constants of binding and unbinding of monomer subunits. Because of the geometry of the polymer end and the monomer-monomer  interactions, the growth properties of two-stranded polymers depend only on three parameters, namely, the rate constants of attaching or dissociating from the leading protofilament and the energy of lateral interactions.

The exact full dynamic description of the growth of two-stranded polymers, that accounts for all possible configurations, has been compared with a set of $n$-layer approximate models that consider only the most relevant polymer configurations. It was shown that the approximate approach is successful because it captures the main features of full dynamic  kinetic diagram. In addition, the approximate description becomes better for larger lateral interactions between the monomer subunits. It has been concluded that $n$-layer approximations might be viewed as  a series expansion of the full dynamic description of polymer growth dynamics. It implies that the approximate approach can be used to describe the growth dynamics of rigid biopolymers with many protofilaments, like microtubules or intermediate filaments.

The full dynamic chemical kinetic model of the growth of two-stranded polymers has been applied to analyze the experimental observations on single actin filaments growth. Using the kinetic rate parameters and the realistic estimate of the lateral interactions  extracted from bulk chemical kinetic measurements of actin filaments, we calculated the overall dispersion in the length fluctuations of single actin filaments. The obtained value of the dispersion was approximately 10 times smaller than the experimentally observed.\cite{fujiwara02}  The difference is  significant and it  implies that other processes, not accounted by current theoretical analysis, contribute to the dispersion of the single actin filaments. It was argued  that this discrepancy is due to the fact that the hydrolysis in the polymer molecule is not accounted in our theoretical approach. 

In addition, we also discussed the concentration dependence of dispersion. Our theoretical calculations suggest that the dispersion of actin filaments depends weakly non-linearly at low concentrations of free monomers, and it approaches the linear dependence at large concentrations. It will be very important to measure the concentration dependence experimentally since it will give a valuable information on the mechanisms of growth  and  it will provide a direct check of the validity of our theoretical picture. 

In a future, we plan to investigate the effect of hydrolysis of the monomers, associated with ATP or its analogs,  on the growth of biopolymers consisting of $N$ parallel rigid protofilaments. For actin filaments ($N=2$) it seems reasonable to extend the current chemical kinetic model, however for biopolymers with larger number of protofilaments, such as microtubules and intermediate filaments, the coupling of hydrolysis with $n$-layer approximate approach, probably, is the most realistic approach.

\section*{Acknowledgments}

The authors would like to acknowledge the support from the Welch Foundation (grant  C-1559), the Alfred P. Sloan foundation (grant  BR-4418)  and the U.S. National Science Foundation (grant CHE-0237105). The authors also are grateful to M.E. Fisher for valuable discussions and encouragements.

\section*{Appendix A: Calculations for full dynamic model of the growth of attached two-stranded polymers}

\setcounter{equation}{0}

\renewcommand{\theequation}{\mbox{A\arabic{equation}}}

Let us introduce the probabilities $P(l,l - k;t)$ and $P(l - k,l + 1;t)$ of finding the two-stranded polymer in the configurations  $(l,l - k)$ and  $(l - k,l + 1)$ respectively  at time $t$. Here  $l, k = 0, 1,...$ and  the two parameters  in brackets correspond to the number of subunits in the first and second protofilaments respectively. These probabilities (at $k=0$) satisfy the following master equations,
\begin{eqnarray}\label{me1}
\frac{dP(l,l;t)}{{dt}} = & u_{1 - \delta } P(l - 1,l;t) + w_\delta  P(l,l + 1;t) + u_1 P(l,l - 1;t) + w_0 P(l + 1,l;t) & \nonumber \\
 &    - (u_\delta + w_{1 - \delta } + u_0  + w_1 )P(l,l;t),  & 
\end{eqnarray}
\begin{eqnarray}\label{me2}
 \frac{dP(l,l + 1;t)}{dt} = & u_\delta  P(l,l;t) + w_{1 - \delta } P(l + 1,l + 1;t) + u_1 P(l - 1,l + 1;t) + w_0 P(l,l + 2;t) & \nonumber \\
 & - (u_{1 - \delta }  + w_\delta   + u_1  + w_0 )P(l,l + 1;t). &  
\end{eqnarray}
These equations  describe  a set of special states on the main pathway in a  chemical kinetic scheme, see Fig. 2a.  For $k \neq 0$ we have
\begin{eqnarray}\label{me3}
 \frac{dP(l,l - k;t)}{dt}  = & u_0 P(l - 1,l - k;t) + w_1 P(l,l + 1 - k;t) + u_1 P(l,l - 1 - k;t) & \nonumber \\
& + w_0 P(l + 1,l - k;t) - (u_0  + w_0  + u_1  + w_1 )P(l,l - k;t),& 
\end{eqnarray}
and
\begin{eqnarray}\label{me4}
\frac{dP(l - k,l + 1;t)}{dt} = & u_0 P(l - k,l)  + w_1 P(l + 1 - k,l + 1;t)  + u_1 P(l - 1 - k,l + 1;t) \nonumber \\
& + w_0 P(l - k,l + 2;t) - (u_0  + w_0  + u_1  + w_1 )P(l - k,l + 1;t). &
\end{eqnarray} 
The conservation of probability leads to  
\begin{equation}
\sum\limits_{l = 0}^{ + \infty } {\left( {\sum\limits_{k = 0}^{ + \infty } {P(l,l - k;t) + \sum\limits_{k = 0}^{ + \infty } {P(l - k,l + 1;t)} } } \right)}  = 1,
\end{equation}
at all times.

Following the idea of  Derrida,\cite{derrida83} we define four sets of auxiliary functions ($k = 0,1,...$),
\begin{eqnarray}
& &  B_{k,0} (t)= \sum\limits_{l = 0}^{ + \infty } {P(l,l - k;t)}  \\
& &  C_{k,0} (t)= \sum\limits_{l = 0}^{ + \infty } {(l + \delta ) P(l,l - k;t)} \\
& &  B_{k,1} (t)= \sum\limits_{l = 0}^{ + \infty } {P(l - k,l + 1;t)}  \\
& &  C_{k,1} (t)= \sum\limits_{l = 0}^{ + \infty } {(l + 1) P(l - k,l + 1;t)} 
\end{eqnarray} 
where $ \delta  = a/d$. Note that the conservation of probability gives us
\begin{equation}\label{norm}
\sum\limits_{k = 0}^{ + \infty } {\sum\limits_{i = 0}^1 {B_{k,i} (t)} }  = 1.
\end{equation}
Then from master equations (\ref{me1}), (\ref{me2}), (\ref{me3}) and  (\ref{me4}) we derive  for $k = 0$
\begin{eqnarray}\label{Be1}
  \frac{dB_{0,0} (t)}{dt} = &(u_{1 - \delta }  + w_\delta  )B_{0,1} (t) + (u_1  + w_0 )B_{1,0} (t) - (u_\delta   + w_{1 - \delta }  + u_0  + w_1 )B_{0,0} (t), & \nonumber \\
  \frac{dB_{0,1} (t)}{dt} = &(u_\delta   + w_{1 - \delta } )B_{0,0} (t) + (u_1  + w_0 )B_{1,1} (t) - (u_{1 - \delta }  + w_\delta   + u_0  + w_1 )B_{0,1} (t), & 
\end{eqnarray} 
while for $k \neq 0$ ($i = 0,1$) it is given by
\begin{equation}\label{Be2}
\frac{dB_{k,i} (t)}{dt} = (u_0  + w_1 )B_{k - 1,i} (t) + (u_1  + w_0 )B_{k + 1,i} (t) - (u_0  + w_0  + u_1  + w_1 )B_{k,i} (t).
\end{equation}

Similar arguments can be used to describe functions $C_{k,0} (t)$  and $C_{k,1} (t)$. Specifically, for $k = 0$ we obtain
\begin{eqnarray}\label{Ce1}
\frac{dC_{0,0} (t)}{dt} = &(u_{1 - \delta }  + w_\delta  )C_{0,1} (t)  + (u_1  + w_0 )C_{1,0} (t) - (u_\delta   + w_{1 - \delta }  + u_0  + w_1 )C_{0,0} (t) & \nonumber \\
& + [\delta u_{1 - \delta }  - (1 - \delta )w_\delta  ]B_{0,1} (t) - w_0 B_{1,0} (t), &
\end{eqnarray}
\begin{eqnarray}\label{Ce2}
\frac{dC_{0,1} (t)}{dt} = &(u_\delta   + w_{1 - \delta } )C_{0,0} (t)  + (u_1  + w_0 )C_{1,1} (t) - (u_{1 - \delta }  + w_\delta   + u_0  + w_1 )C_{0,1} (t) & \\
& + [(1 - \delta )u_\delta   - \delta w_{1 - \delta } ]B_{0,0} (t) - w_0 B_{1,1} (t). &
\end{eqnarray} 
For $k \neq 0$ ($i = 0,1$) the expressions are
\begin{eqnarray}\label{Ce3}
\frac{dC_{k,i} (t)}{dt}= &(u_0  + w_{1} )C_{k - 1,i} (t) + (u_1  + w_0 )C_{k + 1,i} (t)  - (u_0  + w_0  + u_1  + w_1 )C_{k,i} (t)  & \nonumber  \\
& + u_0 B_{k - 1,i} (t) - w_0 B_{k + 1,i} (t).  &
\end{eqnarray}

Again following the Derrida's method,\cite{derrida83} we introduce the ansatz that should be valid at large times $t$, namely,
\begin{equation}\label{ansatz}
B_{k,i} (t) \to b_{k,i} ,{\text{ }}C_{k,i} (t) \to a_{k,i} t + T_{k,i} \hspace{0.5cm} (i = 0,1).
\end{equation}
At steady state $dB_{k,i} (t)/dt = 0$ and Eqs. (\ref{Be1}) and (\ref{Be2})  yield for $k = 0$
\begin{eqnarray}\label{be1}
&  0 = (u_{1 - \delta }  + w_\delta  )b_{0,1}  + (u_1  + w_0 )b_{1,0}  - (u_\delta   + w_{1 - \delta }  + u_0  + w_1 )b_{0,0},  & \nonumber \\
&  0 = (u_\delta   + w_{1 - \delta } )b_{0,0}  + (u_1  + w_0 )b_{1,1}  - (u_{1 - \delta }  + w_\delta   + u_0  + w_1 )b_{0,1},  & 
\end{eqnarray} 
while for $k \neq 0$ ($i = 0,1$) we obtain
\begin{equation}\label{be2}
0 = (u_0  + w_1 )b_{k - 1,i}  + (u_1  + w_0 )b_{k + 1,i}  - (u_0  + w_0  + u_1  + w_1 )b_{k,i}. 
\end{equation}

The solutions of Eqs. (\ref{be1}) and (\ref{be2}) can be written in the following form
\begin{equation}
b_{k,i}  = \frac{{1 - \beta }}{{1 + \alpha }}\alpha ^i \beta ^k,  \hspace{0.5cm} \mbox{for  }(i = 0,1),
\end{equation}
where  $k = 0,1,..$, and
\begin{equation}
\alpha  = \frac{{u_\delta   + w_{1 - \delta } }}{{u_{1-\delta}   + w_\delta  }}; \hspace{0.5cm} \beta  = \frac{{u_0  + w_1 }}{{u_1  + w_0 }}.
\end{equation}

To determine the coefficients $a_{k,i}$ and $T_{k,i}$ from  Eq. (\ref{ansatz}), the ansatz for the functions $C_{k,i} $ is substituted into the asymptotic expressions (\ref{Ce1}),  (\ref{Ce2}) and (\ref{Ce3}), yielding for $k = 0$,
\begin{eqnarray}\label{ae1}
&   0 = (u_{1 - \delta }  + w_\delta  )a_{0,1}  + (u_1  + w_0 )a_{1,0}  - (u_\delta   + w_{1 - \delta }  + u_0  + w_1 )a_{0,0}, & \nonumber \\
&  0 = (u_\delta   + w_{1 - \delta } )a_{0,0}  + (u_1  + w_0 )a_{1,1}  - (u_{1 - \delta }  + w_\delta   + u_0  + w_1 )a_{0,1}.  &
\end{eqnarray} 
At the same time,  for $k \neq 0$ ($i = 0,1$) we obtain
\begin{equation}\label{ae2}
0 = (u_0  + w_1 )a_{k - 1,i}  + (u_1  + w_0 )a_{k + 1,i} - (u_0  + w_0  + u_1  + w_1 )a_{k,i}.
\end{equation}
The coefficients $T_{k,i}$ satisfy the following equations (for $k = 0$)
\begin{eqnarray}\label{te1}
a_{0,0}  = &(u_{1 - \delta }  + w_\delta  )T_{0,1}  + (u_1  + w_0 )T_{1,0}  - (u_\delta   + w_{1 - \delta }  + u_0  + w_1 )T_{0,0} & \nonumber \\
&  + [\delta u_{1 - \delta }  - (1 - \delta )w_\delta  ]b_{0,1}  - w_0 b_{1,0}, &
\end{eqnarray} 
and 
\begin{eqnarray}\label{te2}
a_{0,1}  = & (u_\delta   + w_{1 - \delta } )T_{0,0} + (u_1  + w_0 )T_{1,1}  - (u_{1 - \delta }  + w_\delta   + u_0  + w_1 )T_{0,1}  \nonumber \\
& + [(1 - \delta )u_\delta   - \delta w_{1 - \delta } ]b_{0,0}  - w_0 b_{1,1}.  
\end{eqnarray} 
For $k \neq 0$ ($i = 0,1$) we have
\begin{equation}\label{te3}
a_{k,i}  = (u_0  + w_1 )T_{k - 1,i}  + (u_1  + w_0 )T_{k + 1,i}  - (u_0  + w_0  + u_1  + w_1 )T_{k,i}  + u_0 b_{k - 1,i}  - w_0 b_{k + 1,i}. 
\end{equation}

Comparing Eqs.(\ref{be1}) and (\ref{be2}) with expressions (\ref{ae1}) and (\ref{ae2}), we conclude that
\begin{equation}\label{eq_a}
a_{k,i}  = Ab_{k,i}, \hspace{0.5cm} (i = 0,1),
\end{equation}
with the constant $A$. This constant can be calculated  by summing  over the left and right sides in Eq. (\ref{eq_a}) and recalling the  normalization condition (\ref{norm}). The summation over all  $a_{k,i}$ in Eqs. (\ref{ae1}) and (\ref{ae2}) produces
\begin{equation}
A = \sum\limits_{k = 0}^{ + \infty } {\sum\limits_{i = 0}^1 {a_{k,i} } }  = [(1 - \delta )u_\delta   - \delta w_{1 - \delta } ]b_{0,0}  + [\delta u_{1 - \delta }  - (1 - \delta )w_\delta  ]b_{0,1}  + u_0  - w_0 (1 - b_{0,0}  - b_{0,1} ).
\end{equation}
Thus  we have $A = A_0(1 - \beta)  + A_1$ , where
\begin{equation}
 A_0 = [(1 - \delta )u_\delta   - \delta w_{1 - \delta } ]b_{0,0}  + [\delta u_{1 - \delta }  - (1 - \delta )w_\delta  ]b_{0,1}  = \frac{u_\delta  u_{1 - \delta }  - w_\delta  w_{1 - \delta} } {u_\delta   + w_\delta   + u_{1 - \delta }  + w_{1 - \delta } },
\end{equation}
and
\begin{equation}
A_1  = u_0  - w_0 \beta.   
\end{equation} 
Note that $A_0$ does not depend explicitly on $\delta$.

To determine the coefficients $T_{k,i}$ , we define  for all $k$ the following function
\begin{eqnarray}
&   T_k  \equiv T_{k,0}  + T_{k,1};  &  \\
&   a_k  \equiv a_{k,0}  + a_{k,1};  & \\
&   b_k  \equiv b_{k,0}  + b_{k,1}. &  
\end{eqnarray} 
Then we define  
\begin{equation}
y_k  \equiv (u_1  + w_0 )T_k  - (u_0  + w_1 )T_{k - 1}. 
\end{equation}
Now Eqs. (\ref{te1}), (\ref{te2}) and (\ref{te3}) can be rewritten as
\begin{eqnarray}
&  y_0  - y_{ - 1}  = a_0  + w_0 b_1  - A_0(1 - \beta), & \\
&   y_k  - y_{k - 1}  = a_k  - u_0 b_{k - 1}  + w_0 b_{k + 1},   
\end{eqnarray} 
with $y_{-1} \equiv 0$ and  $k = 1,2,...$. 
The solutions for these equations are given by
\begin{equation}
  y_k  = A_0(1 - \beta) \left( {\sum\limits_{j = 0}^k {b_j }  - 1} \right) + u_0 b_k 
\end{equation}
Summing up $a_{k,0}$  or $a_{k,1}$ separately for all $k$, we obtain the relationship between the parameters $T_{0,1}$ and $T_{0,0}$, i.e.,
\begin{equation}
T_{0,1}  = \alpha T_{0,0}  + \frac{b_{0,0} [(1 - \delta )u_\delta   - \delta w_{1 - \delta } ] - \alpha A_0 (1 - \beta)/ (1 + \alpha )}{u_{1 - \delta }  + w_\delta },
\end{equation}
and for sum of all $T_k$ we have
\begin{equation}
\sum\limits_{k = 0}^\infty  {T_k }  = T_0  + \frac{{u_0  -  A_0 \beta }}{{u_1  + w_0 }}.
\end{equation}

It is now possible to calculate explicitly the mean growth velocity, $V$, and dispersion, $D$, at steady-state conditions. The average length of the polymer is given by
\begin{eqnarray}\label{eq_l}
 < l(t) >  &  = & d\left( {\sum\limits_{k = 0}^{ + \infty } {\sum\limits_{l = 0}^{ + \infty } {(l + \delta ) P(l,l - k;t)} }  + \sum\limits_{k = 0}^{ + \infty } {\sum\limits_{l = 0}^{ + \infty } {(l + 1) P(l - k,l + 1;t)} } }  \right)  \nonumber \\
  & = & d\sum\limits_{k = 0}^{ + \infty } {[C_{k,0} (t) + C_{k,1} (t)]}.  
\end{eqnarray} 
Then, using Eq. (\ref{eq_a}), we obtain for the velocity
\begin{equation}
V = \mathop {\lim }\limits_{t \to \infty } \frac{d}{{dt}} < l(t) >  = dA\left( {\sum\limits_{k = 0}^{ + \infty } {b_{k,0} }  + \sum\limits_{k = 0}^{ + \infty } {b_{k,1} } } \right) = dA.
\end{equation}

A similar approach can be used to derive  the expression for  dispersion. We start from
\begin{equation}
< l^2 (t) >  =  d^2 \left( {\sum\limits_{k = 0}^{ + \infty } {\sum\limits_{l = 0}^{ + \infty } {(l + \delta )^2  P(l,l - k;t)} }  + \sum\limits_{k = 0}^{ + \infty } {\sum\limits_{l = 0}^{ + \infty } {(l + 1)^2  P(l - k,l + 1;t)} } } \right).
\end{equation}
Then, using master equations (\ref{me1}), (\ref{me2}), (\ref{me3}) and  (\ref{me4}), it can be shown that
\begin{eqnarray}\label{eq_l1}
   \mathop{\lim } \limits_{t \to \infty } \frac{d}{dt} <l^2 (t)> & = d^2  \left\{ [(1 - \delta )u_\delta   - \delta w_{1 - \delta } ] C_{0,0} + [\delta u_{1 - \delta }- (1 - \delta )w_\delta  ] C_{0,1}  \right.  & \nonumber \\
  &   + [(1 - \delta )^2 u_\delta   + \delta ^2 w_{1 - \delta } ] b_{0,0}  + [\delta ^2 u_{1 - \delta }  + (1 - \delta )^2 w_\delta  ] b_{0,1} + u_0 & \nonumber \\
 & \left.  + w_0 (1 - b_{0,0}  - b_{0,1} )  + u_0 \sum\limits_{k = 0}^{ + \infty } {[C_{k,0}  + } C_{k,1} ] + w_0 \sum\limits_{k = 1}^{ + \infty } {[C_{k,0}  + } C_{k,1} ]\right\}. &   \end{eqnarray}
Also, the following expression can be derived using Eq. (\ref{eq_l})
\begin{equation}\label{eq_l2}
\mathop {\lim }\limits_{t \to \infty } \frac{d}{{dt}} \left(<l(t)>^2\right)  =2 d^2 A\sum\limits_{k = 0}^{ + \infty } {[C_{k,0}  + C_{k,1} ]}. 
\end{equation}
The formal expression for dispersion is given by
\begin{equation}
D=\frac{1}{2}\mathop {\lim }\limits_{t \to \infty } \frac{d}{{dt}} \left(<l(t)^{2}>-<l(t)>^2 \right). 
\end{equation}
Then substituting into this expression Eqs. (\ref{eq_l1}) and (\ref{eq_l2}) we obtain
\begin{eqnarray}
  D =  \frac{d^2 }{2} & \left\{ [(1 - \delta )u_\delta   - \delta w_{1 - \delta } ]T_{0,0}  + [\delta u_{1 - \delta }  - (1 - \delta )w_\delta  ] T_{0,1}  + 
\frac{1}{2}[(1 - \delta )^2 u_\delta   + \delta ^2 w_{1 - \delta } ] b_{0,0} \right.  & \nonumber \\
 & +\frac{1}{2}[\delta ^2 u_{1 - \delta }  + (1 - \delta )^2 w_\delta  ] b_{0,1}  + \frac{1}{2}u_0  + \frac{1}{2}w_0 (1 - b_{0,0}  - b_{0,1} ) +  w_0 (T_{0,0}  + T_{0,1} )& \nonumber \\
& \left. + (u_0  - w_0  - A)\sum\limits_{k = 0}^{ + \infty } {[T_{k,0}  + } T_{k,1} ] \right \}.  &
\end{eqnarray}
Finally, after some algebraic transformations, we derive the final expression for the dispersion, $D = D_0  + D_1$, which is  given in Eqs. (\ref{dif1}), (\ref{dif2}) and  (\ref{dif3}) in Sec. II. Note that a constant  $T_{0,0}$ cancels out  in the final equation.

\newpage

\noindent {\bf Figure Captions:} \\\\

\noindent Fig. 1 \quad  Different configurations of the growing two-stranded polymer molecule. The molecule is attached at the left end and it can grow only from the right end. The size of the monomer subunit is $d$, while $a$ is a shift between the parallel protofilaments. The rates and labels for different configurations are explained in the text.

\vspace{5mm}

\noindent Fig. 2 \quad  Chemical kinetic schemes for models of the growth of two-stranded attached polymers: a)  full dynamic description; b)  two-layer approximate model; c) one-layer approximate model.

\vspace{5mm}

\noindent Fig. 3 \quad Comparison of the exact dynamic properties of polymer growth  calculated in the full dynamic description with the approximate results from one-layer and two-layer models for different geometries. a) Ratio of exact and approximate mean growth velocities as a function of lateral interactions. b)  Ratio of exact and approximate dispersions as a function of lateral interactions.

\vspace{5mm}

\noindent Fig. 4 \quad Dispersion as a function of free actin monomers for growing single actin filaments.

\newpage

\begin{figure}[ht]
\begin{center}
\vskip 1.5in
\unitlength 1in
\begin{picture}(4.0,4.0)
\resizebox{3.375in}{2.00in}{\includegraphics{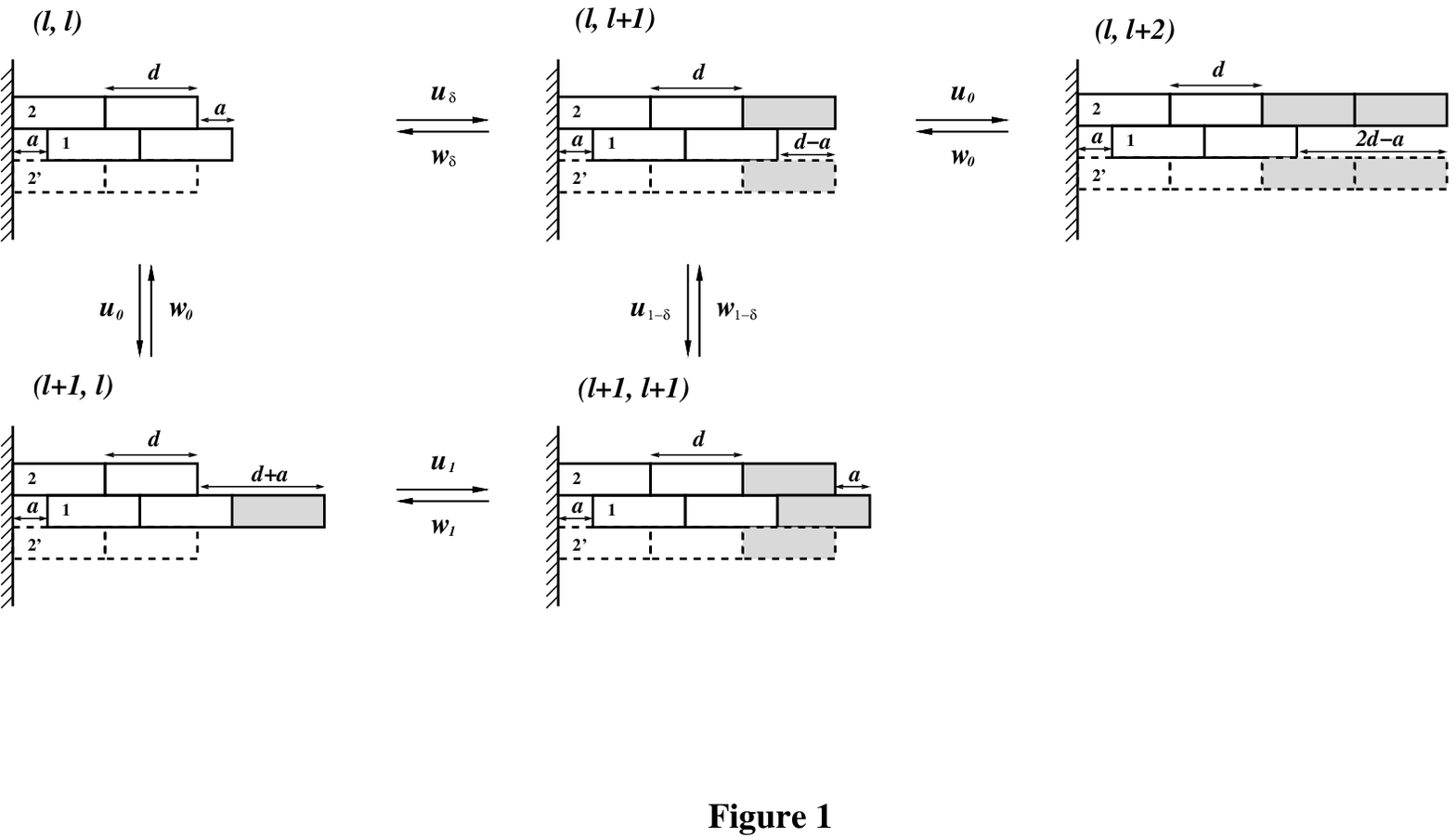}}
\end{picture}
\vskip 3in
 \begin{Large}  \end{Large}
\end{center}
\vskip 3in
\end{figure}

\newpage

\begin{figure}[ht]
\begin{center}
\vskip 1.5in
\unitlength 1in
\begin{picture}(4.0,4.0)
\resizebox{3.375in}{2.71in}{\includegraphics{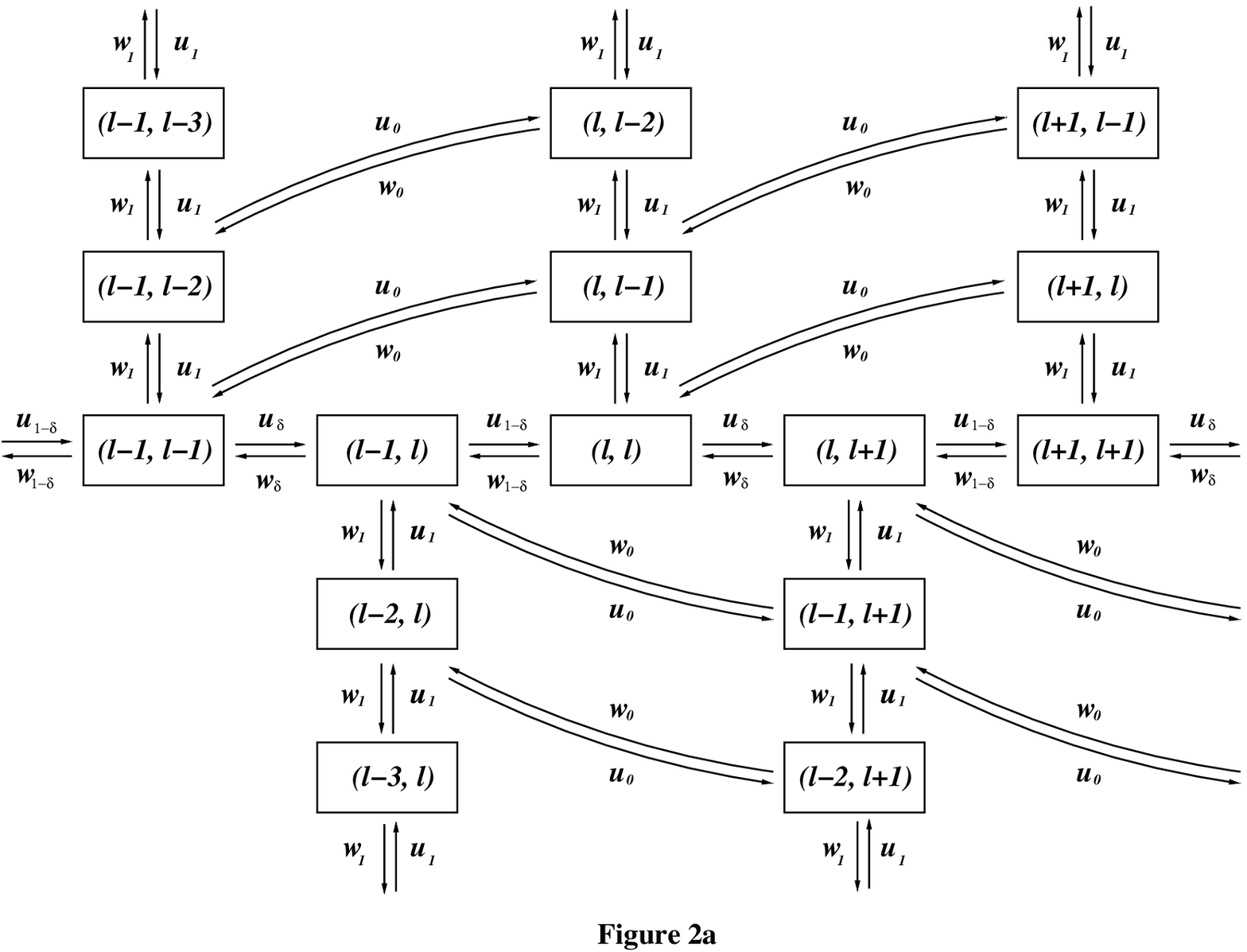}}
\end{picture}
\vskip 3in
 \begin{Large}  \end{Large}
\end{center}
\vskip 3in
\end{figure}

\newpage

\begin{figure}[ht]
\begin{center}
\vskip 1.5in
\unitlength 1in
\begin{picture}(4.0,4.0)
\resizebox{3.375in}{1.34in}{\includegraphics{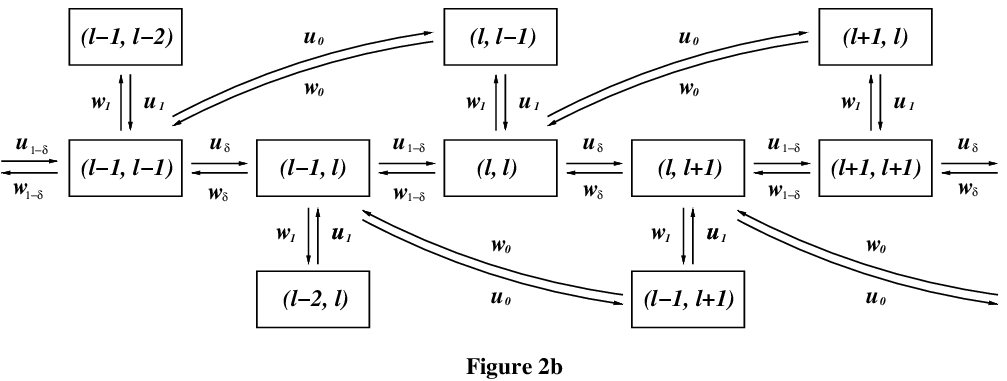}}
\end{picture}
\vskip 3in
 \begin{Large}  \end{Large}
\end{center}
\vskip 3in
\end{figure}

\newpage

\begin{figure}[ht]
\begin{center}
\vskip 1.5in
\unitlength 1in
\begin{picture}(4.0,4.0)
\resizebox{3.375in}{0.53in}{\includegraphics{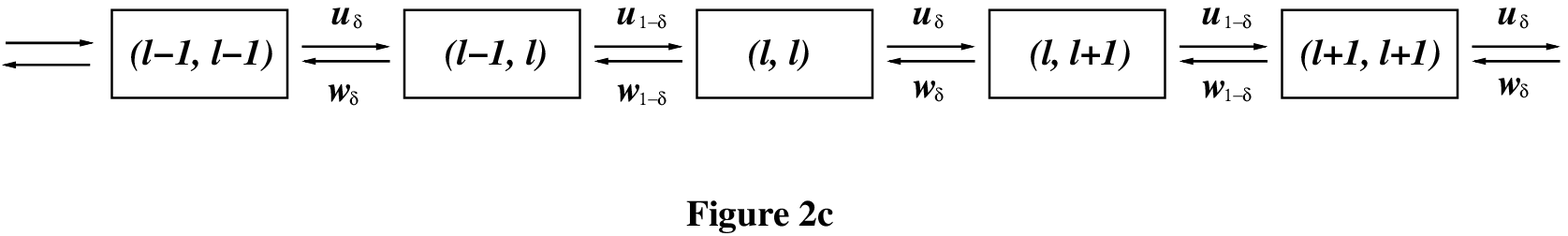}}
\end{picture}
\vskip 3in
 \begin{Large}  \end{Large}
\end{center}
\vskip 3in
\end{figure}

\newpage

\begin{figure}[ht]
\begin{center}
\vskip 1.5in
\unitlength 1in
\begin{picture}(4.0,4.0)
\resizebox{3.375in}{3.375in}{\includegraphics{Figure3a.actin.eps}}
\end{picture}
\vskip 3in
 \begin{Large}  \end{Large}
\end{center}
\vskip 3in
\end{figure}

\newpage

\begin{figure}[ht]
\begin{center}
\vskip 1.5in
\unitlength 1in
\begin{picture}(4.0,4.0)
\resizebox{3.375in}{3.375in}{\includegraphics{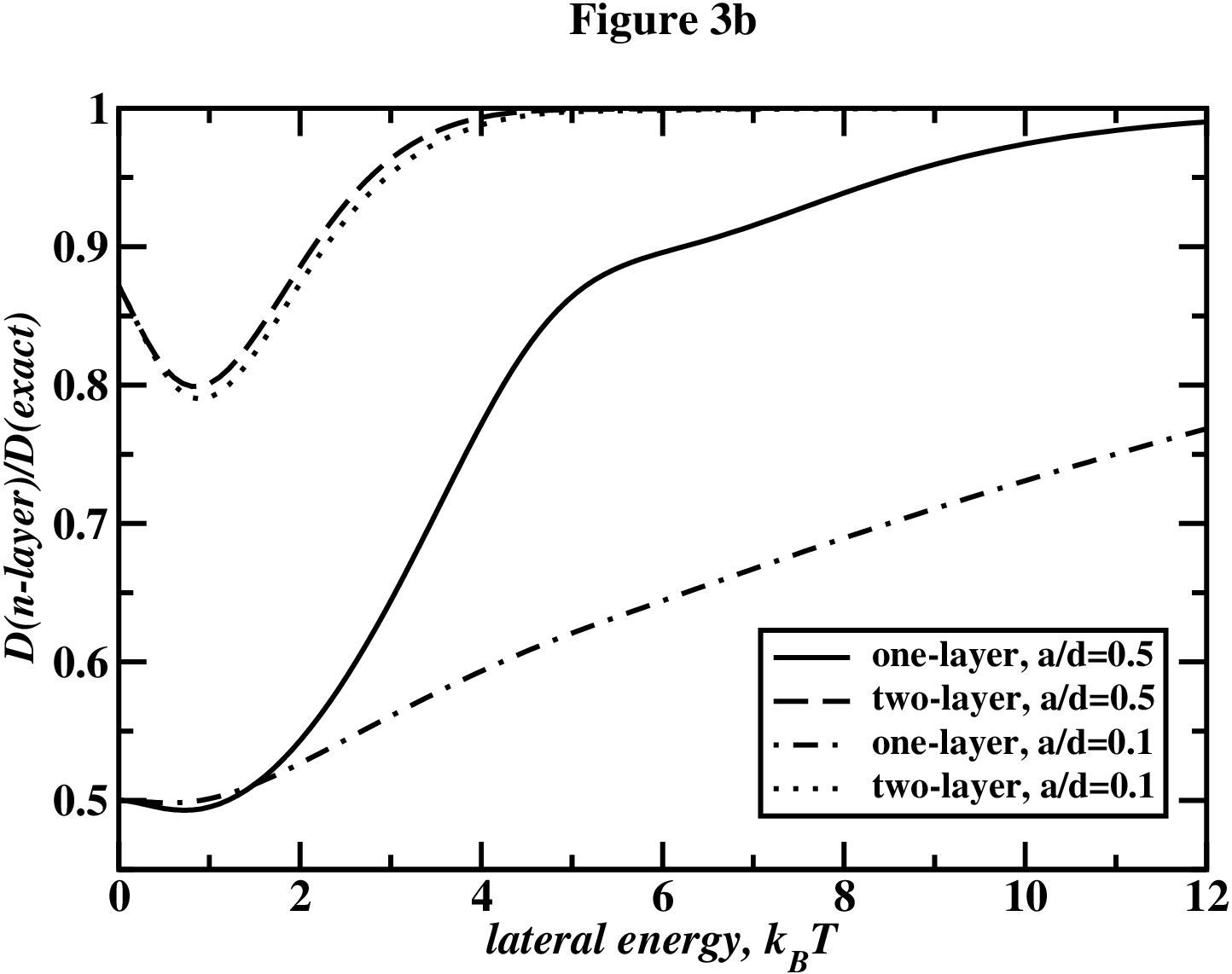}}
\end{picture}
\vskip 3in
 \begin{Large}  \end{Large}
\end{center}
\vskip 3in
\end{figure}

\newpage

\begin{figure}[ht]
\begin{center}
\vskip 1.5in
\unitlength 1in
\begin{picture}(4.0,4.0)
\resizebox{3.375in}{3.375in}{\includegraphics{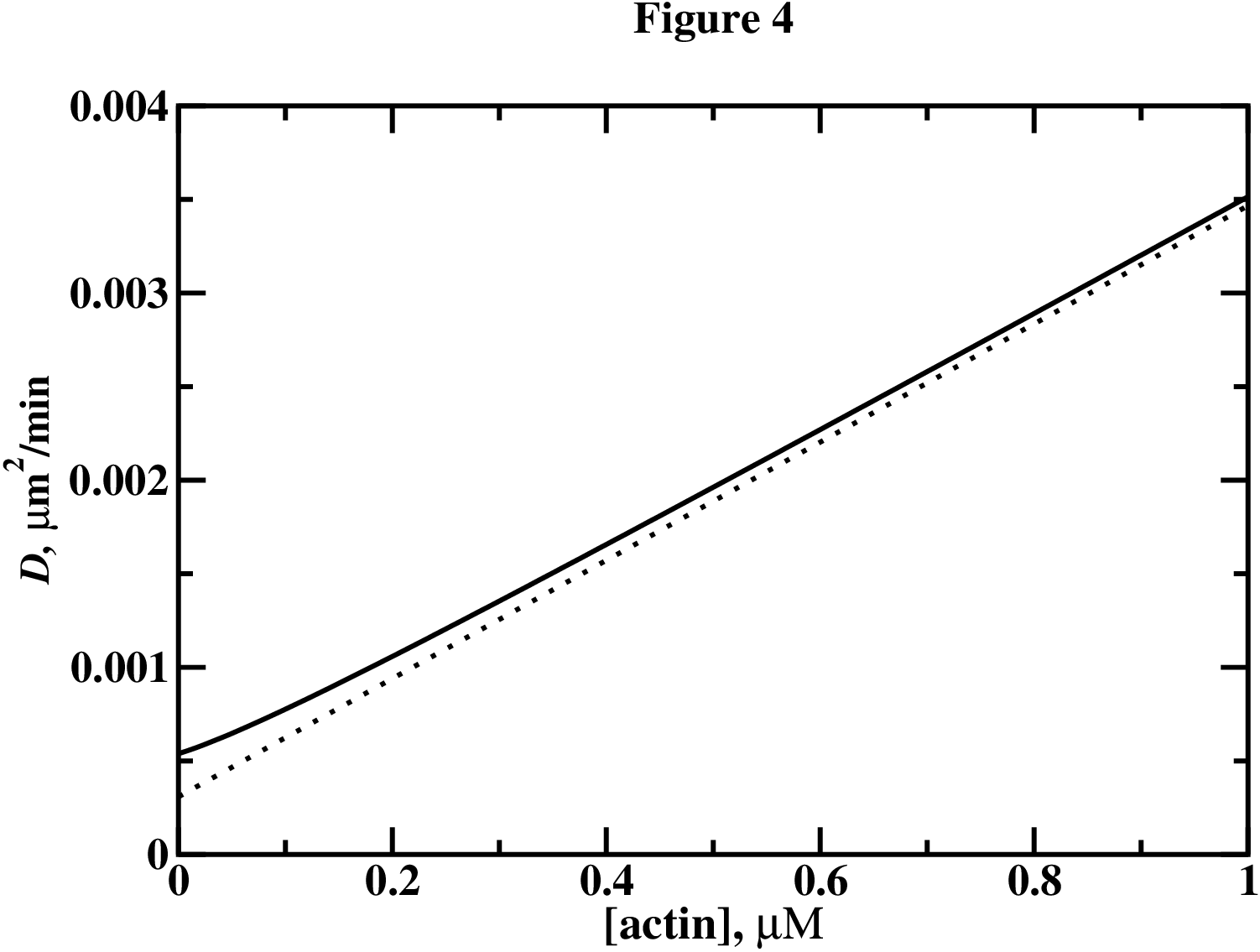}}
\end{picture}
\vskip 3in
 \begin{Large}  \end{Large}
\end{center}
\vskip 3in
\end{figure}

\end{document}